\documentclass[a4paper,11pt]{article}
\usepackage{lipsum}
\usepackage[english]{babel}
\usepackage{chemfig}
\usepackage[utf8]{inputenc}
\usepackage{amsmath}
\usepackage{graphicx}
\usepackage[colorinlistoftodos]{todonotes}
\usepackage[version=3]{mhchem}
\usepackage{chemfig}
\usepackage{cancel}
\usepackage{fancyhdr}
\usepackage{caption}
\usepackage{subcaption}
\usepackage[labelfont=bf]{caption} 
\usepackage[letterpaper,top=2cm,bottom=2cm,left=3cm,right=3cm,marginparwidth=1.75cm]{geometry}
\usepackage[colorlinks=true, allcolors=blue]{hyperref}
\date{}
\usepackage[affil-it]{authblk} 
\usepackage[section]{placeins}

\makeatletter
\renewcommand{\maketitle}{\bgroup\setlength{\parindent}{0cm}
\begin{flushleft}
  \LARGE{\textbf{Ecological transition for the gas mixtures of the MRPC cosmic ray telescopes of the EEE Project}\newline
  }

  \@author
\end{flushleft}\egroup
}

\title{\textbf{Ecological transition for the gas mixtures of the MRPC cosmic ray telescopes of the EEE Project}}

\def\correspondingauthor{\footnote{Correspondig author: cripoli@unisa.it}}
\author[a,b]{C.~Ripoli \correspondingauthor{}}
\author[p,c]{M.~Abbrescia}
\author[d,e]{C.~Avanzini}
\author[e,d]{L.~Baldini}
\author[f]{R.~Baldini~Ferroli}
\author[e,d]{G.~Batignani}
\author[g]{M.~Battaglieri}
\author[h,i]{S.~Boi}
\author[d,e]{E.~Bossini}
\author[j]{F.~Carnesecchi}
\author[a]{D.~Cavazza}
\author[i]{C.~Cical\`{o}}
\author[k,o]{L.~Cifarelli}
\author[l]{F.~Coccetti}
\author[m]{E.~Coccia}
\author[n]{A.~Corvaglia}
\author[a,b]{D.~De~Gruttola}
\author[a,b]{S.~De~Pasquale}
\author[q]{L.~Galante}
\author[l,o]{M.~Garbini}
\author[l,r]{I.~Gnesi}
\author[t,g]{S.~Grazzi}
\author[o,j]{D.~Hatzifotiadou}
\author[u,w]{P.~La~Rocca}
\author[v]{Z.~Liu}
\author[x]{L.~Lombardo}
\author[t,w]{G.~Mandaglio}
\author[o]{A.~Margotti}
\author[y]{G.~Maron}
\author[c]{M.~N.~Mazziotta}
\author[h,i]{A.~Mulliri}
\author[o]{R.~Nania}
\author[o]{F.~Noferini}
\author[z]{F.~Nozzoli}
\author[k,o]{F.~Palmonari}
\author[za,n]{M.~Panareo}
\author[n]{M.~P.~Panetta}
\author[zb,d]{R.~Paoletti}
\author[x]{M.~Parvis}
\author[zc]{C.~Pellegrino}
\author[g]{L.~Perasso}
\author[o]{O.~Pinazza}
\author[zd]{C.~Pinto}
\author[l,f]{S.~Pisano}
\author[u,w]{F.~Riggi}
\author[ze]{G.~Righini}
\author[c]{M.~Rizzi}
\author[k,o]{G.~Sartorelli}
\author[a]{E.~Scapparone}
\author[zf,r]{M.~Schioppa}
\author[k,a]{G.~Scioli}
\author[zb,d]{A.~Scribano}
\author[o]{M.~Selvi}
\author[zg,g]{M.~Taiuti}
\author[d]{G.~Terreni}
\author[t,w]{A.~Trifir\`{o}}
\author[t,w]{M.~Trimarchi}
\author[zc]{C.~Vistoli}
\author[zh]{L.~Votano}
\author[j,v]{M.~C.~S.~Williams}
\author[l,k,o,j,v]{A.~Zichichi}
\author[v,j]{R.~Zuyeuski}

\affil[a]{Dipartimento di Fisica "E. R. Caianiello", Universit\`{a} di Salerno, via Giovanni Paolo II, 132, 84084 Fisciano (SA), Italy}

\affil[b]{INFN, Gruppo Collegato di Salerno, Complesso Universitario di Monte S. Angelo ed. 6 via Cintia, 80126, Napoli, Italy}

\affil[c]{INFN, Sezione di Bari, via Orabona 4, 70126 Bari, Italy}

\affil[d]{INFN, Sezione di Pisa, largo Bruno Pontecorvo 3, 56127 Pisa, Italy}

\affil[e]{Dipartimento di Fisica "E. Fermi", Universit\`{a} di Pisa, largo Bruno Pontecorvo 3, 56127 Pisa, Italy}

\affil[f]{INFN, Laboratori Nazionali di Frascati, via Enrico Fermi 54, 00044 Frascati (RM), Italy}

\affil[g]{INFN, Sezione di Genova, via Dodecaneso, 33, 16146 Genova, Italy}

\affil[h]{Dipartimento di Fisica, Universit\`{a} di Cagliari, S.P. Monserrato-Sestu Km 0,700, 09042 Monserrato (CA), Italy}

\affil[i]{INFN, Sezione di Cagliari, Complesso Universitario di Monserrato, S.P. per Sestu – Km 0,700, 09042 Monserrato (CA), Italy}

\affil[j]{i, Esplanade des Particules 1, 1211 Geneva 23, Switzerland Geneva}

\affil[k]{Dipartimento di Fisica e Astronomia "A. Righi", Universit\`{a} di Bologna, viale Carlo Berti PIchat 6/2, 40127 Bologna}

\affil[l]{Museo Storico della Fisica e Centro Studi e Ricerche "E. Fermi", via Panisperna 89/a, 00184 Roma, Italy}

\affil[m]{Gran Sasso Science Institute, viale Francesco Crispi 7,  67100 L'Aquila, Italy}

\affil[n]{INFN, Sezione di Lecce, via per Arnesano. 73100, Lecce, Italy}

\affil[o]{INFN, Sezione di Bologna, viale Carlo Berti Pichat 6/2, 40127 Bologna}

\affil[p]{Dipartimento di Fisica “M. Merlin” dell’Universit\`{a} e del Politecnico di Bari, via Amendola 173, 70125 Bari, Italy}

\affil[q]{Teaching and Language Lab (q), Politecnico di Torino, corso Duca degli Abruzzi 24, Torino, Italy}

\affil[r]{INFN, Gruppo Collegato di Cosenza, via Pietro Bucci, Rende (Cosenza), Italy}

\affil[t]{Dipartimento di Scienze Matematiche e Informatiche, Scienze Fisiche e Scienze della Terra, Universit\`{a} di Messina, viale Ferdinando Stagno d'Alcontres 31, 98166 Messina (ME), Italy}

\affil[u]{Dipartimento di Fisica e Astronomia "E. Majorana", Universit\`{a} di Catania, via S. Sofia 64, 95123 Catania, Italy}

\affil[v]{ICSC World Laboratory, Geneva, Switzerland }

\affil[w]{INFN, Sezione di Catania, via S. Sofia 64, 95123 Catania, Italy}

\affil[x]{Dipartimento di Elettronica e Telecomunicazioni, Politecnico di Torino, corso Duca degli Abruzzi 24, Torino, Italy}

\affil[y]{INFN, Laboratori Nazionali di Legnaro, viale dell'Università 2, 35020 Legnaro, Italy}

\affil[z]{INFN, Trento Institute for Fundamental Physics and Applications, via Sommarive, 14, 38123 Povo (TN), Italy}

\affil[za]{Dipartimento di Matematica e Fisica "E. De Giorgi", Universit\`{a} del Salento, via per Arnesano. 73100, Lecce, Italy}

\affil[zb]{Dipartimento di Scienze Fisiche, della Terra e dell’Ambiente, Universit\`{a} di Siena, via Roma 56, 53100 Siena, Italy}

\affil[zc]{INFN-CNAF, viale Carlo Berti PIchat 6/2, 40127 Bologna, Italy}

\affil[zd]{Physik Department, Technische Universitat Munchen, James-Franck-Straße 1, 85748 Garching bei München}

\affil[ze]{CNR, Istituto di Fisica Applicata "Nello Carrara", via Madonna del Piano 10, 50019 Sesto Fiorentino (FI), Italy}

\affil[zf]{Dipartimento di Fisica, Università della Calabria, via Pietro Bucci, Rende (CS), Italy}

\affil[zg]{Dipartimento di Fisica, Università di Genova, via Dodecaneso, 33, 16146 Genova (GE), Italy}

\affil[zh]{INFN, Laboratori Nazionali del Gran Sasso, via G. Acitelli 22, 67100 Assergi (AQ), Italy\newline}

\begin{document}
\maketitle

\begin{abstract}
The Extreme Energy Events (EEE) Collaboration is fully involved in an ecological transition. The use of the standard gas mixture, \ce{C_{2}H_{2}F_{4}}+ \ce{SF_{6}}, has stopped in favor of an alternative green mixture based on \ce{C_{3}H_{2}F_{4}}  with the addition of He or \ce{CO_{2}}.
The choise of these new mixtures is motivated by the significant lower Global Warming Potential (GWP) to reduce the emission of gases potentially contributing to the greenhouse effect. 
The EEE experiment consists of 61 muon telescopes based on Multigap Resistive Plate Chambers (MRPCs), each telescope composed of 3 chambers filled with gas.
Several EEE detectors are today completely fluxed with the new ecological mixture. This contribution will report recent results about the telescope performance obtained from studies with the eco-friendly alternative mixture carried out in the last years.

\end{abstract}

\section{Introduction}
The EEE Project aims to study cosmic rays by detecting  secondary muons on the Earth surface generated by primary cosmic rays interaction in atmosphere. The Project has also with an educational purpose with the direct involvement of Italian high school students and teachers in all the phases of the experiment \cite{garbini2022outreach}.
The EEE network is the largest MRPC-based system consisting of 61 tracking detectors (telescopes). The EEE detectors are installed in Italian high-school buildings and physics laboratories, spanning an area of more than 10$^5$ $km^2$ (Figure 1). 
\begin{figure}[!ht]
\centering
\includegraphics[width=0.30\linewidth]{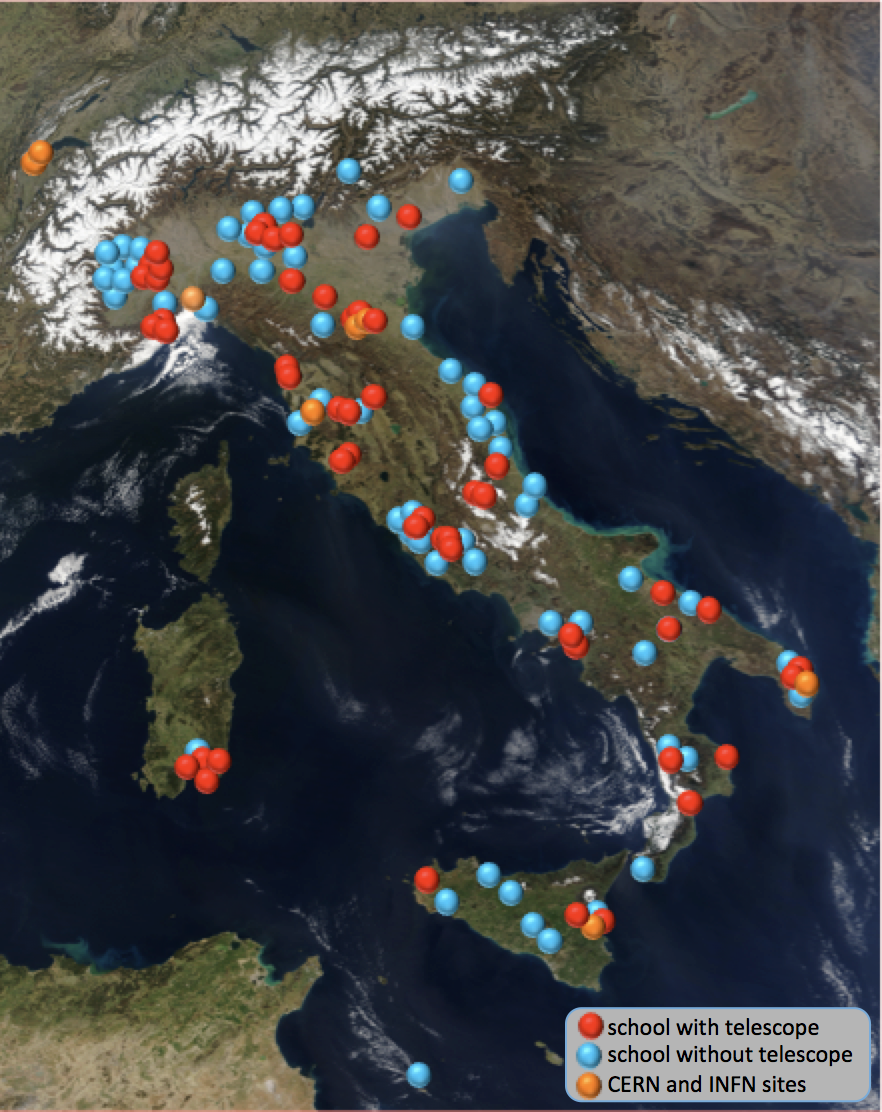}
\caption{\label{fig:EEEitalia_26Feb2019}Map of EEE telescopes and schools participating in the EEE Project. In red and light blue respectively schools with and without the telescope, in orange research centres equipped with telescopes.}
\end{figure}

The telescopes are GPS synchronised for offline analysis on time correlated events. 

The EEE gas detector is composed of 3 large area (0.82 x 1.58)$m^2$ MRPCs operating in avalanche mode, each chamber has six 300 $\mu$m gas gaps (250 $\mu$m in the new chambers built in the EEE upgrade phase since 2017)\cite{abbrescia2018extreme}. 

\begin{figure}[!ht]
\centering
\includegraphics[width=5.0cm]{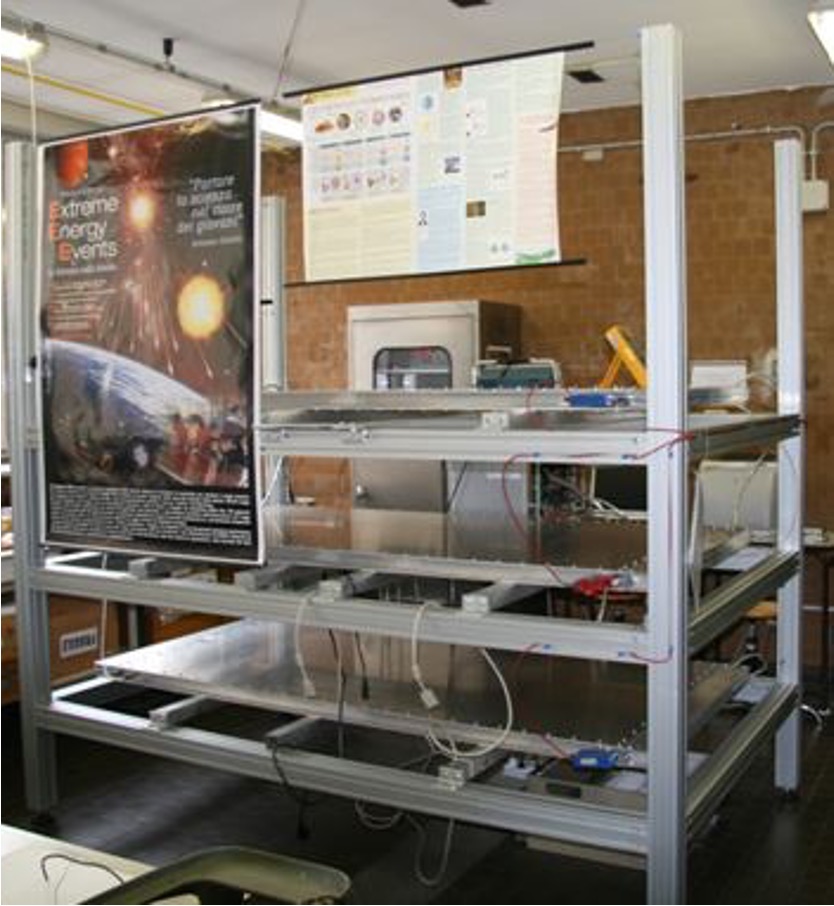}
\includegraphics[width=7.7cm]{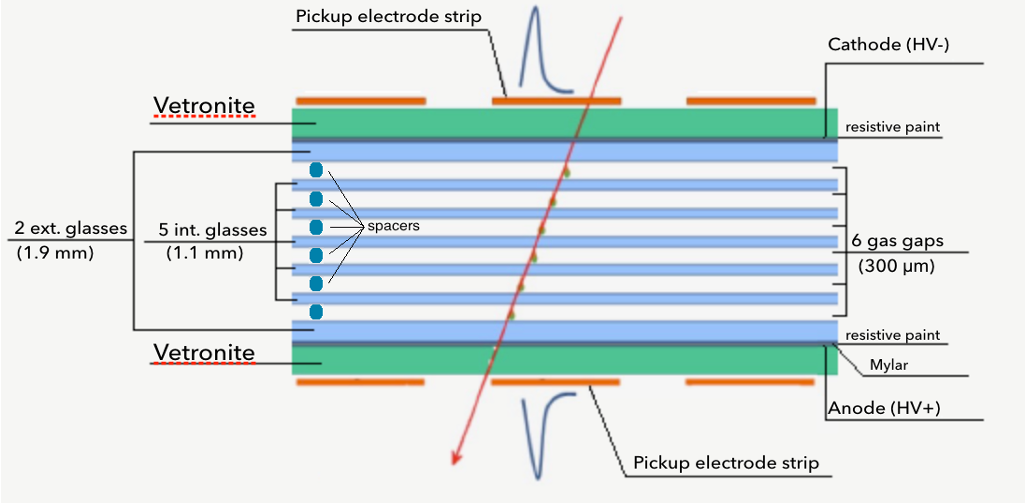}
\caption{EEE telescope (left). Inner structure of a MRPC(right).}
\end{figure}

The standard gas mixture used to flux the chambers consisted of 98\% \ce{C_{2}H_{2}F_{4}} + 2\% \ce{SF_{6}} that are Green House Gases (GHG).
The EEE Collaboration has pursued a path to reduce its GHG emissions searching for an ecological mixture to replace the previous one without affecting the detector performance \cite{ABBRESCIA2023168431, ripoli2022reduction}.
The new mixture has to be compatible with the hardware of the experimental setup and obviously satisfy the safety requirements. Specifically, the mixture has to be characterized by a working point up to $20 kV$ and can only be binary, since each EEE station is equipped with only two flowmeters, which cannot be replaced in the whole array, as the costs would be too high. 
In addition the eco-mixture has to meet a few criteria: compared to the previous one, it needs to have similar performance and a lower GWP that is the the index used to quantify the impact of GHG, each gas has a specific GWP normalized to \ce{{CO_2}} which has a value equal to 1.
The GWP of the EEE standard mixture is a combination of the value of its main component \ce{C_{2}H_{2}F_{4}} (tetrafluoroethane), which is 1430, and of the quencher, \ce{SF_{6}}  (sulfur hexafluoride), equal to 23600; the resulting GWP is 1880, out of the range allowed by the EU regulations banning gases with an index greater than 150. Although research activities exempted from the regulation, the EEE Collaboration decided to start a transition towards a green mixture and the muon telescopes are today fluxed with an eco-friendly gas mixture. 

\section{Eco-friendly mixture }

The search for an eco-friendly gas mixture for the cosmic rays telescopes of the EEE experiment led to replace \ce{C_{2}H_{2}F_{4}} with \ce{C_{3}H_{2}F_{4}} (HFO1234ze - tetrafluoropropene), characterized by a significantly lower GWP, equal to 6. Potentially a mixture made of 100\% of \ce{C_{3}H_{2}F_{4}}  would not require a quencher, however the operating voltage would be too high for the EEE power supply (upper limit 20kV). In order to reduce the operating voltage, a small percentage of Helium is added to the mixture \cite{abbrescia2016eco}. The choice of the eco-friendly mixture was driven by the results of an extensive set of tests \cite{ripoli2022transition}.
Quite a number of mixtures based on \ce{C_{3}H_{2}F_{4}} with different percentages of He have been tested in order to optimize the HV curve. An example is shown in Figure 3, the efficiency test curves of the telescope installed in Pisa INFN Laboratory (named PISA-01)  are shown, comparing the standard mixtures with several percentages of \ce{C_{3}H_{2}F_{4}} and He. The results show that the EEE detectors can be operated with a mixture made of these two gases without any loss in performance \cite{bossini2023studies}.

\begin{figure}[!ht]
\centering
\includegraphics[width=0.60\linewidth]{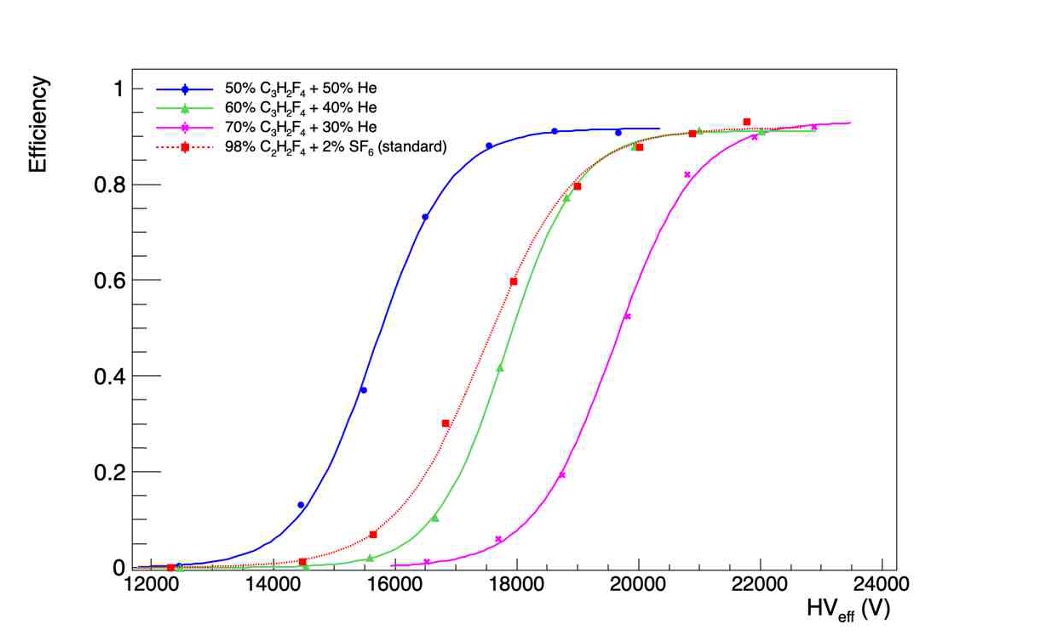}
\caption{\label{fig:newEff}Efficiency curves for mixtures of \ce{C_{3}H_{2}F_{4}} and different percentages of Helium. Dashed red line represents the “standard” mixture, solid lines are used for the \ce{C_{3}H_{2}F_{4}} + He mixtures.
}
\end{figure}
\FloatBarrier

The triple coincidence data acquired with this new eco-friendly mixture shows the expected angular distribution (the zenithal angle of the reconstructed tracks and the angle with respect to the long side of the telescopes) compared with the data collected with the standard mixture \cite{ABBRESCIA2023168431}. Figure 4 show the trigger rate and fraction of reconstructed tracks, identified by a $\chi^2<$ 10, for the SALE-02 EEE telescope with a good rate stability and no degradation in terms of cosmic track reconstruction. 

\begin{figure}[!ht]
\centering
\includegraphics[width=7.7cm]{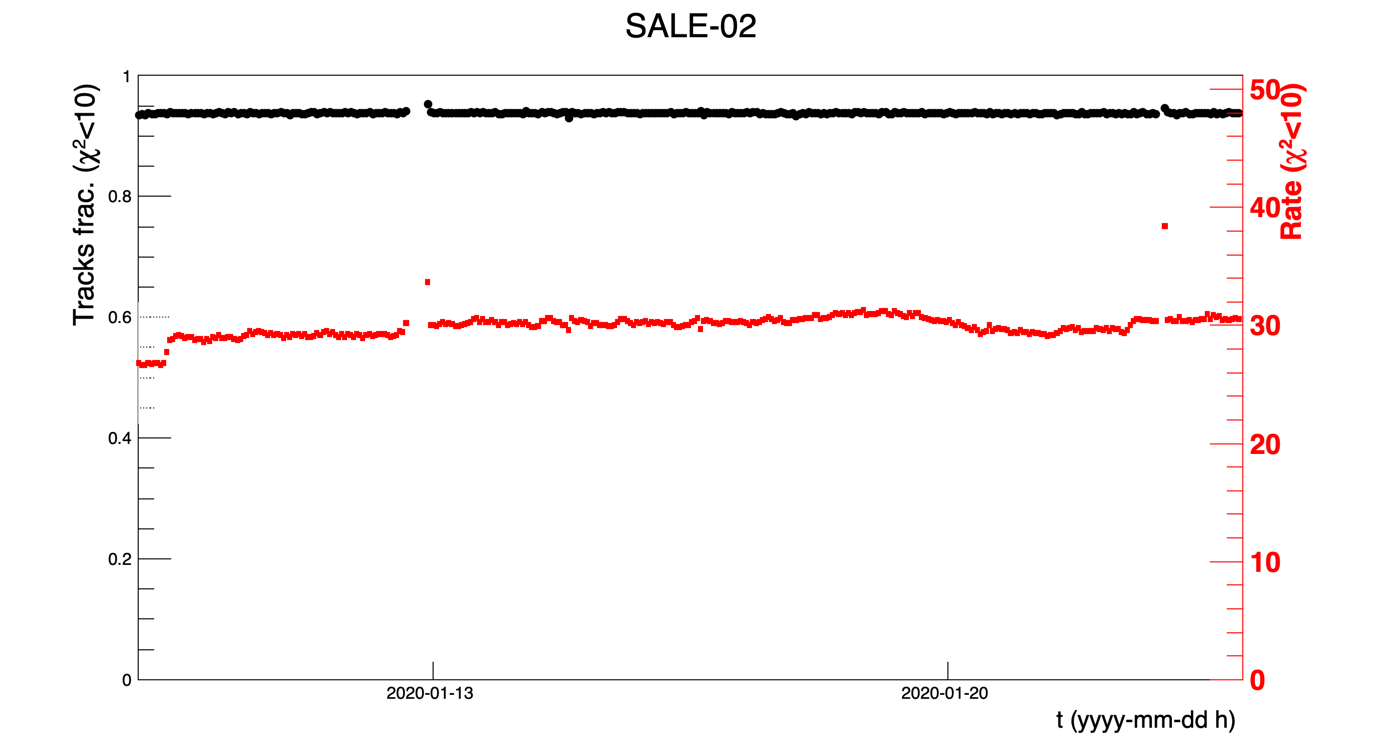}
\includegraphics[width=7.7cm]{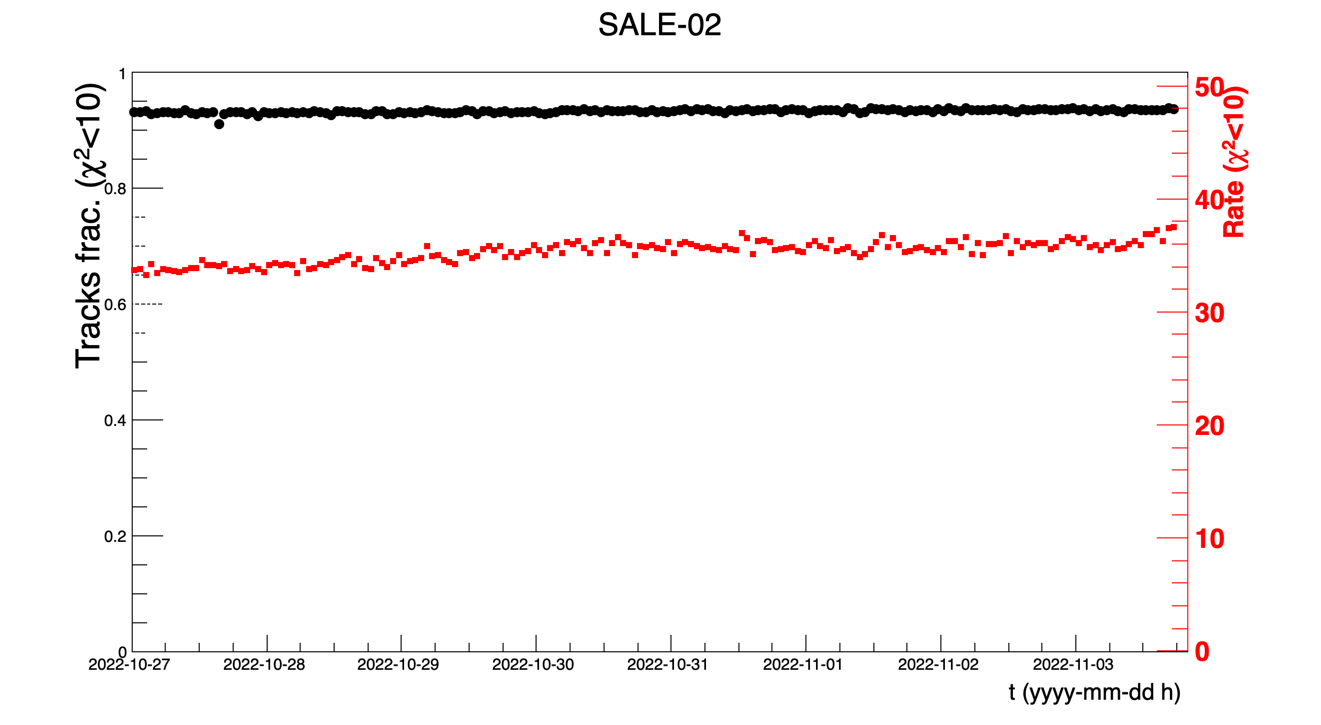}
\caption{Trigger rate in red and fraction of reconstructed tracks in black for the SALE-02 EEE telescope with the standard mixture (left) and with the eco-mixture (right).}
\end{figure}
The change in the mixture does not affect the detector's performance: the trend is approximately constant on the whole data taking period.

The EEE detectors fluxed with the new ecological mixture are: BOLO-02, CAGL-01, CARI-01, CATZ-01, LNLE-01, PISA-01, REND-01, SALE-02, SALE-03. 

\section{Conclusion}
The EEE Collaboration is carrying out an ecological transition to replace the gas mixture used in the muon tracking detectors. Several stations are today in acquisition with the eco-friendly gas mixture made of \ce{C_{3}H_{2}F_{4}}  + He, with satisfactory performance for the physics aims of the experiment and the long term data taking is ongoing. 
The eco-friendly gas mixture currently used guarantees a significant reduction of GWP providing similar performance at the same operating voltage as the standard mixture and no hardware changes are needed.
More telescopes are gradually restarting the data taking with new eco-gas mixture.
Other mixtures, based on \ce{C_{3}H_{2}F_{4}} and \ce{CO_{2}} are under investigation.

\bibstyle{plain}
\bibliography{bibliography.bib}
\end{document}